\documentclass[singlecolumn,a4paper,notitlepage]{revtex4-1}
\usepackage{graphicx}
\newcommand{\deriv}[2]{\frac{\partial #1}{\partial #2}}
\newcommand{\boutxx}{\texttt{BOUT++}}

\begin{document}
\title{The effects of non-uniform drive on plasma filaments}
\author{Brendan Shanahan}
\affiliation{Max-Planck Institut f\"ur Plasmaphysik, Teilinstitut Greifswald, Germany}
\email{brendan.shanahan@ipp.mpg.de}
\author{Ben Dudson and Peter Hill}
\affiliation{York Plasma Institute, Department of Physics, University of York, Heslington, York YO10 5DD, UK}

\begin{abstract}
  Wendelstein 7-X core fueling is primarily achieved through pellet injection.  The trajectory of plasmoids from an ablating pellet is an ongoing research question, which is complicated by the complex magnetic geometry of W7-X; curvature drive varies significantly toroidally, including a change in the drift drive direction.  Here we use the Hermes model in \boutxx~to simulate cold plasma filaments in slab geometries where the magnetic drift drive is non-uniform along the field line.  It is shown that if the field-line-averaged curvature drive is non-zero, a filament will propagate coherently in the direction of average drive.  It is also shown that a non-uniform drive will provide a non-uniform propagation; an effect which is reduced at higher temperatures due to an increased sound speed along the field line.  Finally, simulations with curvature similar to that found in Wendelstein 7-X are performed which indicate a slow radial propagation of plasma filaments despite a highly non-uniform drive profile.
\end{abstract}
\maketitle 
\section{Introduction}
The first experimental campaigns on Wendelstein 7-X have indicated that pellet injection can provide efficient core fueling.  The mechanism by which fast core fueling in W7-X is achieved with ablated pellet clouds is as of yet poorly understood.  Furthermore, poloidally-rotating structures which resemble plasma filaments have been observed near the edge of Wendelstein 7-X plasmas~\cite{Kocsis2017}.  These structures appear to span several toroidal turns.

Given that the curvature drive in Wendelstein 7-X can vary drastically within one toroidal period -- even switching direction -- one could imagine that the opposing curvature drives shear filamentary structures before they are able to propagate.  Since pellet ablation clouds initially have high transport along field lines, this could also affect the transport of pellet clouds.  In this work we look to explore the affects of varying curvature drive on cold filamentary structures using fluid simulations within the \boutxx~\cite{Dudson2009} framework.  

While \boutxx~is capable of simulating complex geometries~\cite{Shanahan2016,ShanahanJPCS2016,Hill2017}, one can often explore complex phenomena by simplifying the problem.  For this reason, isolated filament simulations in slab geometries are often employed~\cite{Walkden2016,Easy2014,Riva2016}.  Here, we look to simplify the effects of non-axisymmetric stellarator geometries by simulating cold plasma filaments in a periodic slab geometries with varying curvature.  This way, it is possible to isolate and investigate the effects of non-uniform drive on filament transport.

\section{Numerical Methods}

\subsection{Hermes plasma model}
While there have been many models implemented in \boutxx, we choose here the Hermes model~\cite{Dudson2017,Leddy2017} for several reasons.  Firstly, a pellet cloud represents a large perturbation in the local density, and therefore a full-profile model which does not use the Boussinesq approximation should be used.  Secondly, a pellet is colder than the surrounding plasma, and therefore a non-isothermal model should be utilized.  Finally, there has been extensive effort in implementing and testing conservative numerical methods in Hermes, which minimizes numerical errors allowing for accurate representation of the analytic model.

A more complete description of the open-source Hermes model can be found in Reference~\cite{Dudson2017}, but the model is repeated here for clarity.  The plasma equations assume cold ions, and evolve the plasma (electron) density $n$, electron thermal pressure $p_e = nT_e$, and vorticity $\omega$, parallel ion momentum $nv_{||i}$ and a form of Ohm's law:
\newpage
\begin{eqnarray}
  \deriv{n}{t} &=& -\nabla\cdot\left(n\mathbf{V}_{E\times B} + n\mathbf{V}_{mag}\right) - \nabla_{||}\left(n_e v_{||e}\right) + \nabla\cdot\left(D_n\nabla_\perp n\right) \nonumber \\
  && + S_n - S \\
  \frac{3}{2}\deriv{p_e}{t} &=& -\nabla\cdot\left(\frac{3}{2}p_e\mathbf{V}_{E\times B} + p_e\frac{5}{2}\mathbf{V}_{mag}\right) - p_e\nabla\cdot\mathbf{V}_{E\times B} \nonumber \\
  && - \frac{5}{2}\nabla_{||}\left(p_e v_{||e}\right) + v_{||e}\partial_{||}p_e  + \nabla_{||}\left(\kappa_{e||}\partial_{||}T_e\right) \nonumber \\
  && + 0.71\nabla_{||}\left(T_e j_{||}\right) - 0.71 j_{||}\partial_{||} T_e + \frac{\nu}{n}j_{||}\left(j_{||} - j_{||0}\right)\nonumber \\
  &&+ \nabla\cdot \left(\frac{3}{2}D_nT_e\nabla_\perp n\right) + \nabla\cdot\left(\chi n\nabla_\perp T_e\right) + S_p - Q\\
  \deriv{\omega}{t} &=&  -\nabla\cdot\left[\frac{1}{2}\left(\omega + \frac{n}{B}\nabla_\perp^2\phi\right)\frac{\mathbf{b}\times\nabla\phi}{B}\right] + \nabla_\perp\cdot\left(\frac{1}{2}\frac{\partial n_i}{\partial t}\frac{1}{B^2}\nabla_\perp\phi\right) \nonumber \\
  &&+ \nabla_{||}j_{||} - \nabla\cdot\left(n \mathbf{V}_{mag}\right)  + \nabla\cdot\left(\mu_i\nabla_\perp\omega\right)  \\
  \frac{\partial}{\partial t}\left(nv_{||i}\right) &=& -\nabla\cdot\left[nv_{||i}\left(\mathbf{V}_{E\times B}+ \mathbf{b}v_{||i}\right)\right] - \partial_{||}p_e \nonumber\\
  && + \nabla\cdot \left(D_nv_{||i}\nabla_\perp n\right) - F \\
  \frac{\partial}{\partial t}\left[\frac{m_e}{m_i}\left(v_{||e}-v_{||i}\right) + \frac{1}{2}\beta_e\psi\right] &=& \nu j_{||}/n_e + \partial_{||}\phi - \frac{1}{n_e}\partial_{||} p_e - 0.71\partial_{||} T_e \label{eq:ohmslaw} \\
  &&+ \frac{m_e}{m_i}\left(\mathbf{V}_{E\times B} + \mathbf{b}v_{||i}\right)\cdot\nabla\left(v_{||i} - v_{||e}\right)
\end{eqnarray}
with E$\times$B and magnetic drifts given by:
\begin{equation}
\mathbf{V}_{E\times B} = \frac{\mathbf{b}\times\nabla \phi}{B} \qquad \mathbf{V}_{mag} = -T_e\nabla\times\frac{\mathbf{b}}{B}
\end{equation}
Due to the large density perturbation, all simulations presented here are performed without the Boussinesq approximation (although this is an option in Hermes), and therefore the vorticity is given by
\begin{equation}
\omega = \nabla\cdot\left(\frac{n}{B^2}\nabla_\perp \phi\right)
\end{equation}
Parallel derivatives are described using the following notation:
\begin{equation}
  \partial_{||} f \equiv \mathbf{b}\cdot\nabla f \qquad \nabla_{||} f \equiv \nabla\cdot\left(\mathbf{b} f\right)
\end{equation}
As the plasma model is cast in conservative form, it conserves particle number, and an energy:
\begin{equation}
E = \int dv \frac{m_in}{2B^2}\left|\nabla\phi\right|^2 + \frac{1}{2}m_inV_{||i}^2 + \frac{3}{2}p_e + \frac{1}{4}\beta_e\left|\nabla_\perp\psi\right|^2 + \frac{m_e}{m_i}\frac{1}{2}\frac{j_{||}^2}{n}
\end{equation}
where the terms correspond to the ion $E\times B$ energy, ion parallel kinetic energy, electron thermal energy, and electromagnetic field energy.  Differential operators are discretised using flux-conservative Finite Volume methods, which are discussed in Reference~\cite{Dudson2017}.


\subsection{Initial Conditions}
\begin{figure}[h]
  \begin{center}
    \begin{minipage}{24pc}
      \begin{center}
        \includegraphics[width=21pc]{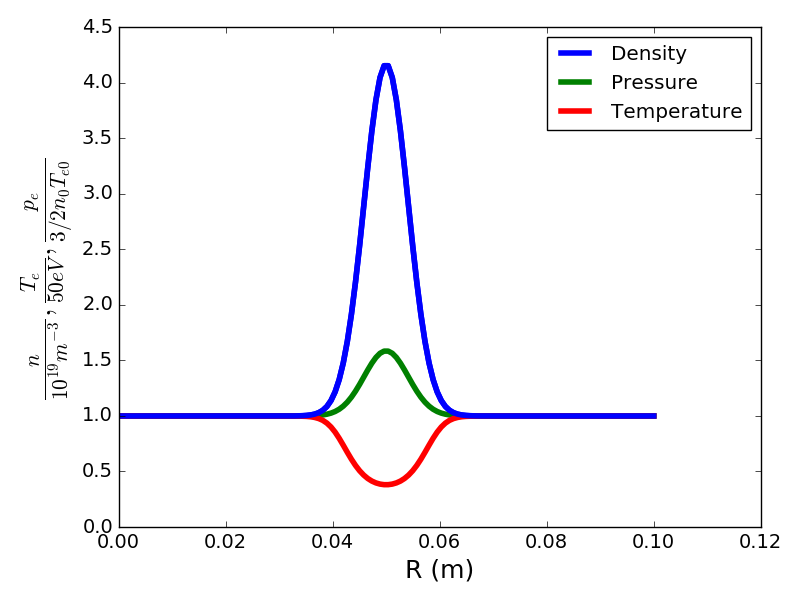}\hspace{2pc}%
        \caption{\label{fig:initialconditions}Initial radial profiles for normalized density (blue), temperature (red) and pressure (green) for cold filament simulations in slab geometries.}
      \end{center}
    \end{minipage}
  \end{center}
\end{figure}

This work seeks to shed light on the evolution of pellet ablation clouds in a complicated magnetic geometry.  For this reason, large, cold filaments were initialized with a positive pressure perturbation.  These filaments are circular in the R-Z plane (x-z in \boutxx~coordinates), and constant along the field line (the \boutxx~y coordinate).  Figure~\ref{fig:initialconditions} illustrates the initial temperature, density and pressure profiles in normalized units for the following simulations.  In these simulations, density was normalized to $n_0 = 1e19 m^{-3}$, temperatures were normalized to $T_{e0} = 50eV$ and pressures were normalized to $p_0 = \frac{3}{2}n_0T_{e0}$.

All simulations presented here were performed with periodic boundary conditions, thereby replicating a small flux tube in the cold field line region of the plasma.  The results are still rather general, as a hotter filament will simply increase the propagation speed, due to the increased pressure perturbation within the filament.

The slab grids had a resolution of 132x16x128 (x,y,z or in real space R, $\parallel$, Z) where the radial and vertical domain have a length of 10cm ($\sim$0.8mm grid spacing), while the parallel domain spans (0,$2\pi$).  

\section{Varying Curvature in a Slab}
\subsection{Non-uniform drive}
\begin{figure}[h]
  \begin{center}
    \begin{minipage}{24pc}
      \begin{center}
        \includegraphics[width=21pc]{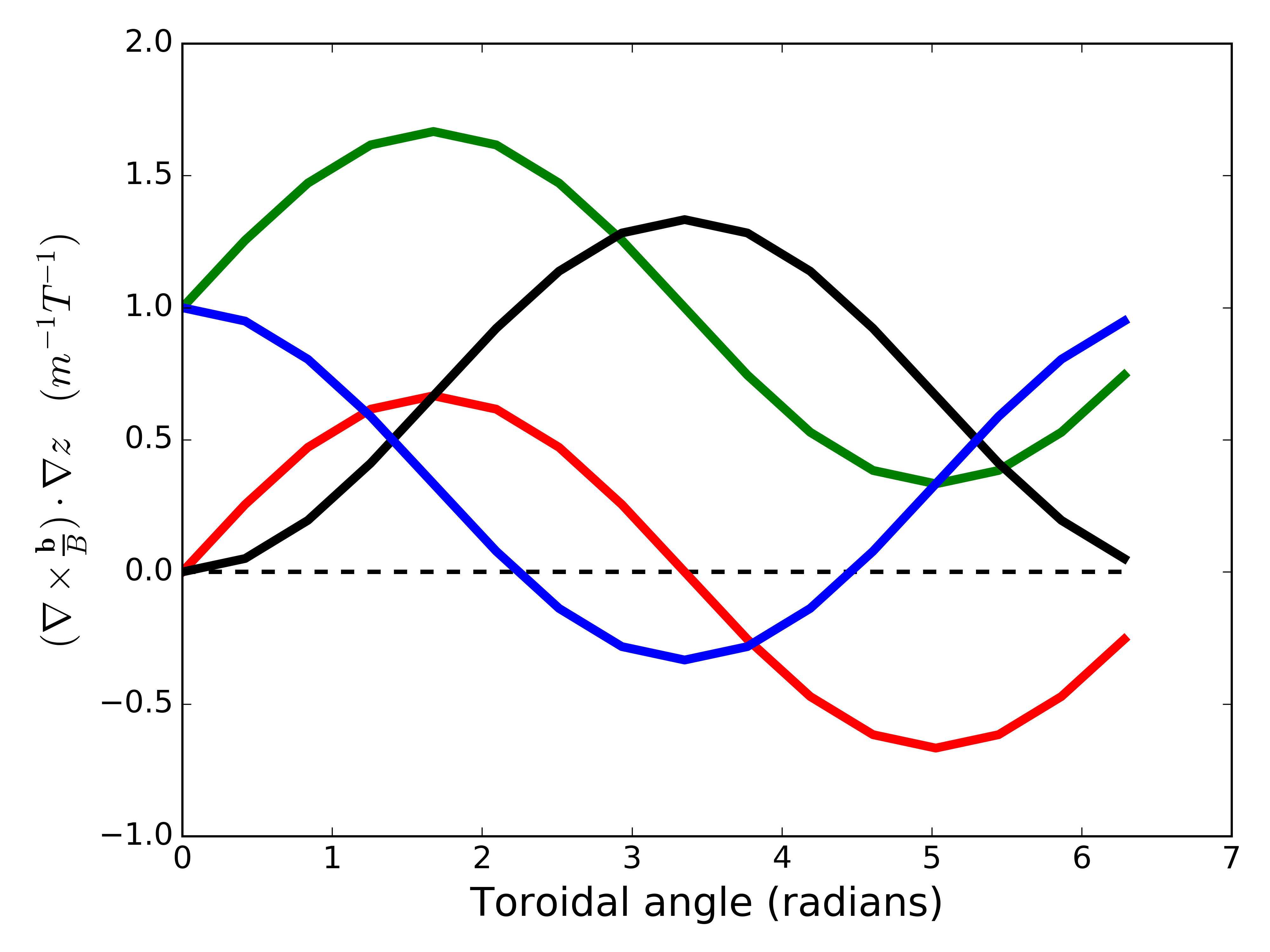}\hspace{2pc}%
        \caption{\label{fig:curv_profiles} Non-uniform curvature drives which were implemented.  A zero-net-curvature drive (red), a non-uniformly positive drive (green), a zero-to-positive drive (black), and a drive which is mostly positive, but dips below zero (blue).}
      \end{center}
    \end{minipage}
  \end{center}
\end{figure}

Examples of the different forms of the filament drive -- in the form of field line curvature -- which are explored here are shown in Figure~\ref{fig:curv_profiles}.  Namely, we explore the effects of a zero-net-curvature drive (red), a drive which is non-uniformly positive (green), a zero-to-positive drive (black) and a mostly positive drive which is briefly negative (blue).  If a pressure gradient which points radially inward is assumed, then a negative field line curvature would correspond to ``good curvature'', which is ballooning stable, and positive drive would be a region of ``bad curvature'', which is ballooning unstable~\cite{Boozer2005}.  Although this is a purely theoretical exercise, the various curvature profiles can be likened to tokamak-like curvature (green), stellarator-like curvature (blue), and two fictional machines: a snaking linear device (red), and a wedge-shaped toroidal pinch (black).   

Filaments with non-uniform drive propagate following the profile of the drive term.  To illustrate this, a quantity $n_{\Delta}$ is introduced, which we define as the difference between the density after 1 microsecond and the initial distribution: $n_{\Delta}(x,y,z) = n(x,y,z,t=1\mu s) - n(x,y,z,t=0)$.

\begin{figure}[h]
  \begin{center}
    \begin{minipage}{24pc}
      \begin{center}
        \includegraphics[width=21pc]{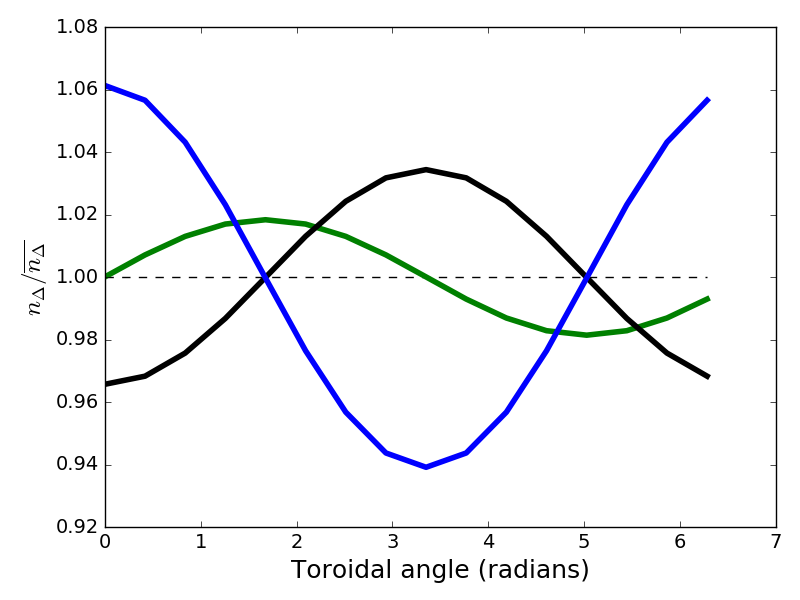}\hspace{2pc}%
        \caption{\label{fig:ndiffcomp}The effects of non-uniform drive; $n_\Delta$ for the cases in Figure~\ref{fig:curv_profiles}, with the exception of the zero-net-drive case, where the filament shears apart.  Colors correspond to those in Figure~\ref{fig:curv_profiles}.}
      \end{center}
    \end{minipage}
  \end{center}
\end{figure}

Although the filaments travel for much longer than 1$\mu s$, a trajectory is already established before the shape of the filaments become less cohesive at later times -- generally with the typical ``mushroom'' shape.  By plotting $n_{\Delta}$ along the peak density at 1$\mu s$, it is possible to visualize the trajectory of the filament while the filament is still cohesive, which facilitates comparison; the higher $n_{\Delta}$, the greater the propagation.  Figure~\ref{fig:ndiffcomp} indicates the $n_{\Delta}$ profiles for the various drive scenarios, excluding the net-zero drive case, where the peak density remains in the middle despite inward- and outward-propagating sections of the filament.

The first result of note is that a curvature drive which changes direction (blue) does not cause the filament to shear apart; rather, the filament propagates radially outward, albeit in a non-uniform fashion.  Only in the case of zero-averaged curvature does the filament shear apart.

\begin{figure}[h]
  \begin{center}
    \begin{minipage}{24pc}
      \begin{center}
        \includegraphics[width=21pc]{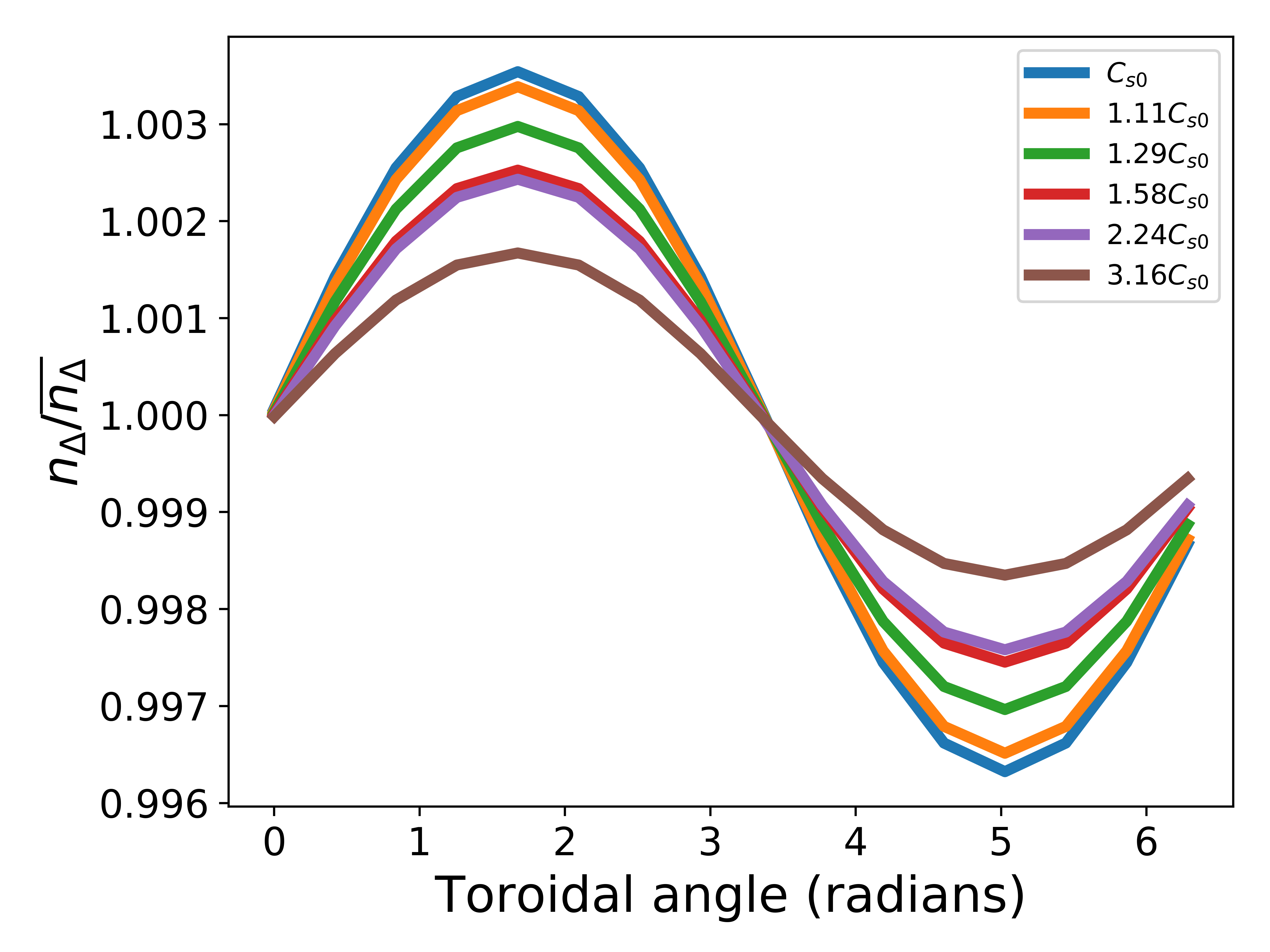}\hspace{2pc}%
        \caption{\label{fig:tscan} $n_\Delta$ relative to the mean for the positive non-uniform drive case with varying sound speed.  A higher temperature (and therefore sound speed) increases the uniformity of the propagation, reducing $n_\Delta / \overline{n_\Delta}$}  
      \end{center}
    \end{minipage}
  \end{center}
\end{figure}

It is also clear from Figure~\ref{fig:ndiffcomp} that the propagation of filaments tends to follow the profile of the responsible drive.  This effect can be reduced, however, by increasing the temperature within the filament -- thereby increasing the sound speed allowing for faster restoration of uniformity.  Figure~\ref{fig:tscan} indicates the $n_{\Delta}$ profiles for cases with varying temperature (and therefore sound speed).  

\subsection{Scaling with average drive}

\begin{figure}[h]
  \begin{center}
    \begin{minipage}{24pc}
      \begin{center}
        \includegraphics[width=21pc]{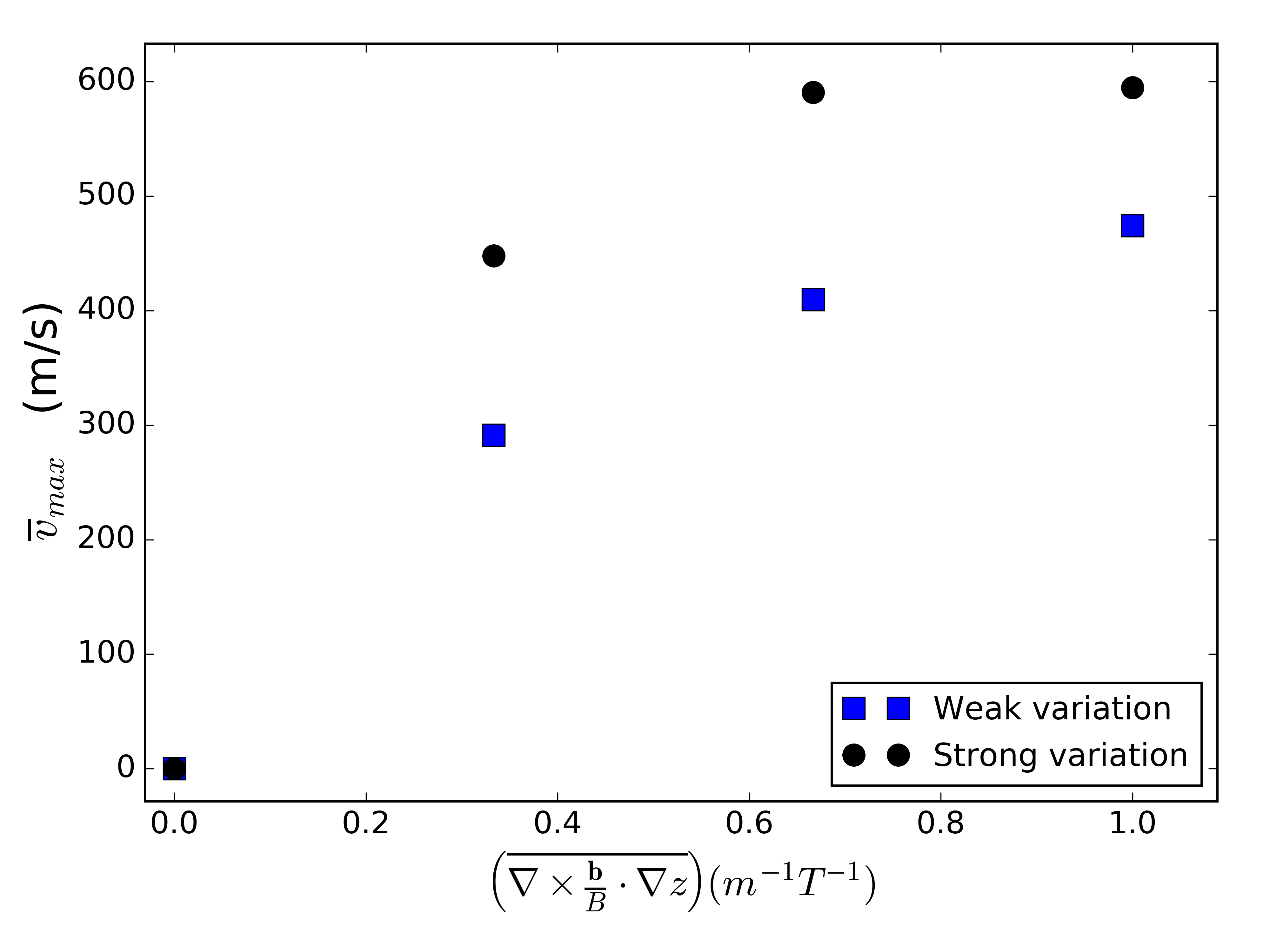}\hspace{2pc}%
        \caption{\label{fig:curv_scaling} The scaling of filament speed with average curvature for two cases; one case where the amplitude of the drive fluctuation is that presented in Figure~\ref{fig:curv_profiles} (blue squares), and one where the amplitude is 2.5 times greater (black circles).  Filament speed scales with average curvature, but the size of the fluctuation is relevant, indicating another mechanism.}
      \end{center}
    \end{minipage}
  \end{center}
\end{figure}

For tokamak filaments, filament propagation is generally considered to scale with the curvature drive (namely $1/R_c$).  In situations with varying curvature drive, this scaling must be adapted.  As an initial investigation, the filament propagation as a function of average curvature drive was plotted in Figure~\ref{fig:curv_scaling}.  To determine if the average curvature is responsible for propagation, the variation in curvature drive was increased by increasing the fluctuation amplitude (but maintaining the average curvature) relative to that shown in Figure~\ref{fig:curv_profiles}.  Figure ~\ref{fig:curv_scaling} indicates these cases with strong variation (black circles) as well as the cases with weaker variation (blue squares).     

While there is a clear increase in propagation speed for increasing average curvature drive, the cases with strong and weak variation exhibit different characteristics -- thereby indicating an additional mechanism for the scaling.

Nevertheless, it can be concluded that in the absence of radial electric fields, any toroidal device should exhibit predominantly radial propagation of plasma filaments, as the curvature must have a nonzero average.  

\subsection{Filaments driven by Wendelstein-7-X-like curvature}
\begin{figure}[h]
  \begin{center}
    \begin{minipage}{24pc}
      \begin{center}
        \includegraphics[width=21pc]{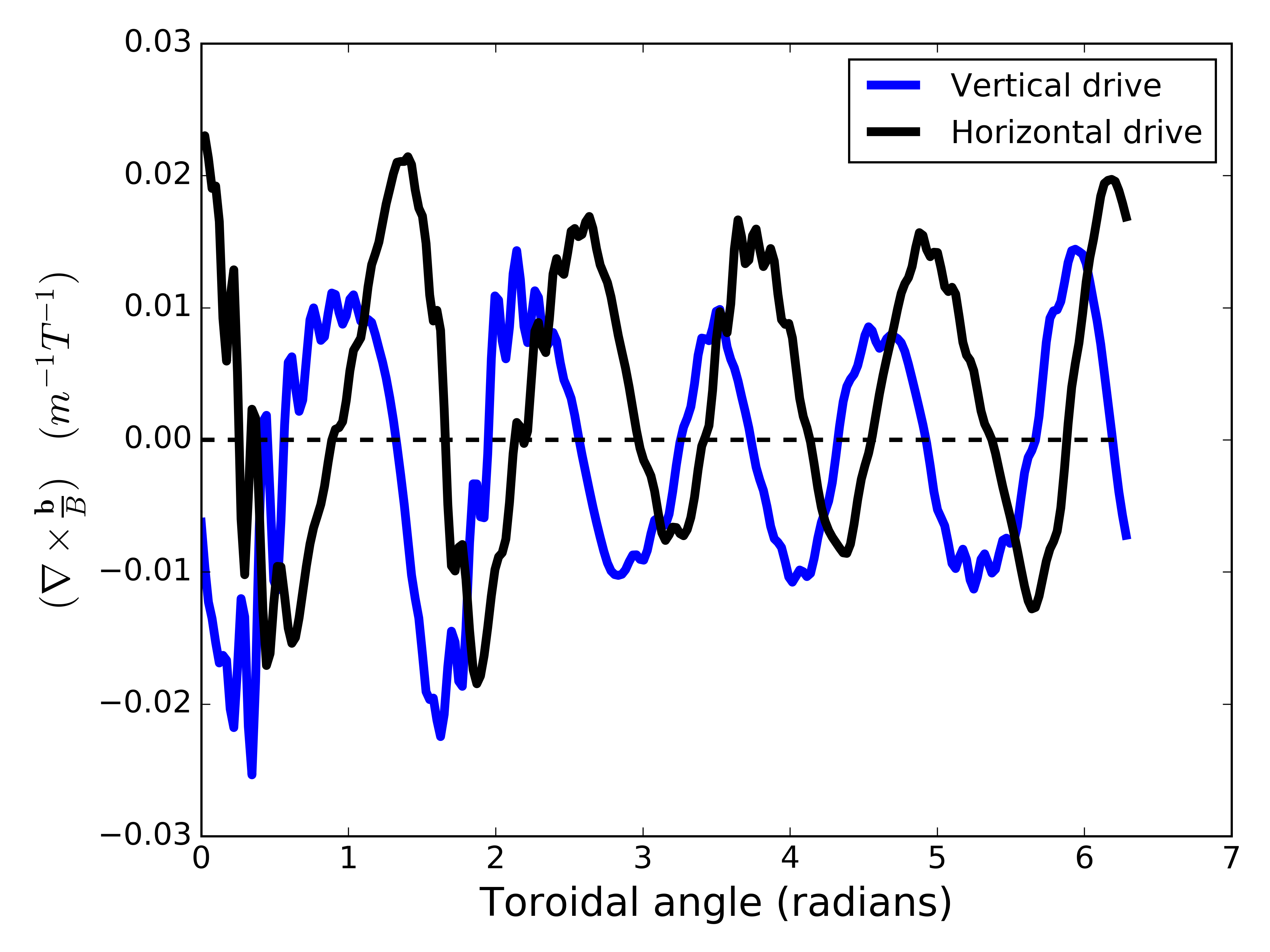}\hspace{2pc}%
        \caption{\label{fig:w7x_curvature} Radial (horizontal, black) and vertical (blue) curvature for a field line in Wendelstein 7-X spanning a full toroidal period and which lies along the inboard pellet injection trajectory.}
      \end{center}
    \end{minipage}
  \end{center}
\end{figure}

The methods in the previous section can be applied to Wendelstein 7-X flux tubes.  To determine the curvature drive, a field line which lies along the inboard pellet injection trajectory was traced for an entire toroidal turn using the Wendelstein 7-X web services~\cite{Bozhenkov2013}, and the vertical and radial curvature drives were calculated, as shown in Figure~\ref{fig:w7x_curvature}.

While the field line characteristics shown in Figure~\ref{fig:w7x_curvature} will change the longer a field line is followed, a single toroidal period ($2\pi$) was chosen here for simplicity.  As W7-X is a low shear device, an assumption is made that the curvature drive does not vary drastically and a periodic parallel boundary condition can be used.

\begin{figure}[h]
  \begin{center}
    \begin{minipage}{34pc}
      \begin{minipage}{16pc}
        \includegraphics[width=15pc]{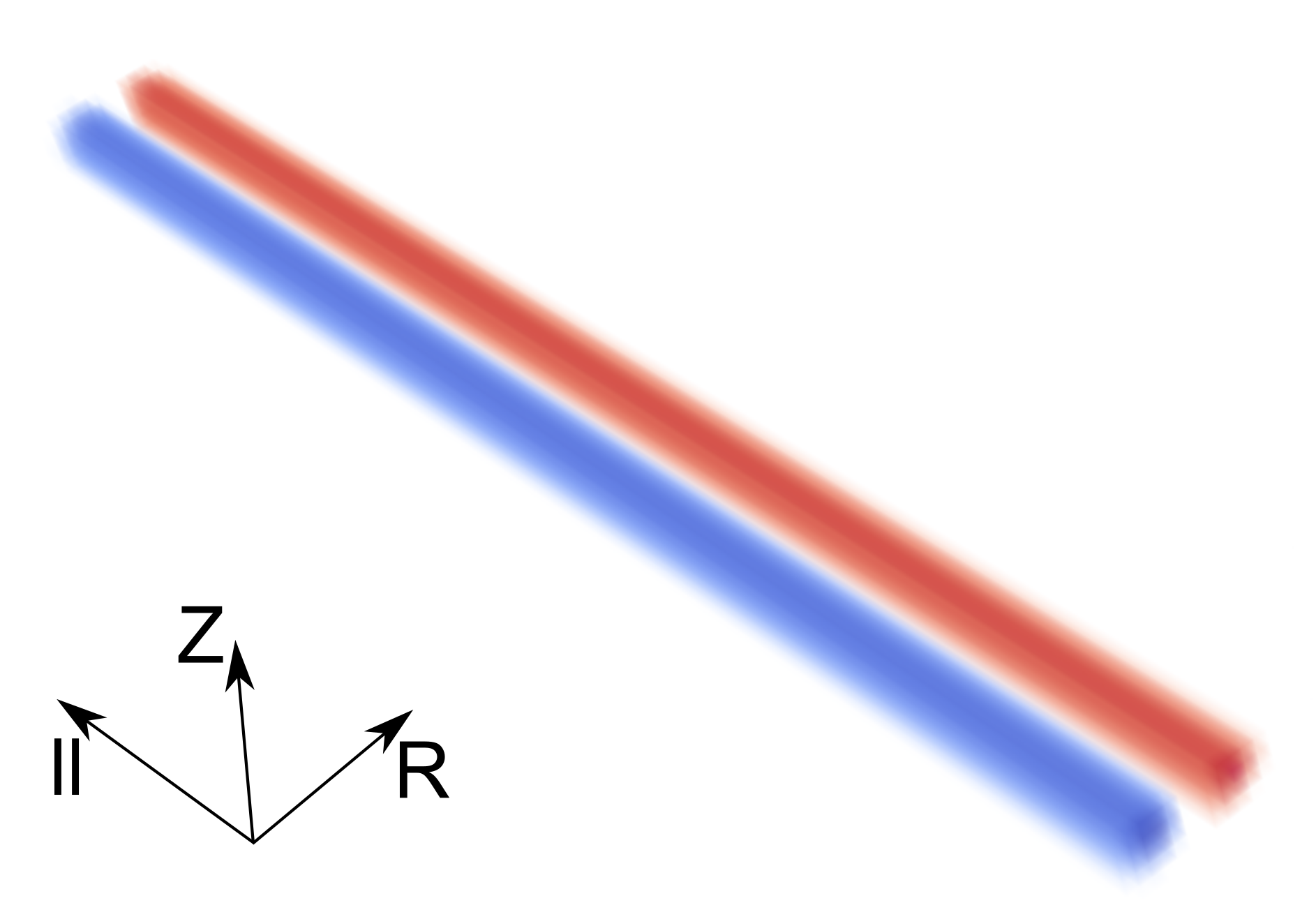}
        \caption{\label{fig:ndiff_w7x} $n_\Delta$ for the Wendelstein 7-X flux tube filament, indicating primarily radial transport, as $n_\Delta$ is positive (red) at the outboard side and negative (blue) at the inboard side.}
      \end{minipage}\hspace{2pc}%
      \begin{minipage}{16pc}
        \includegraphics[width=15pc]{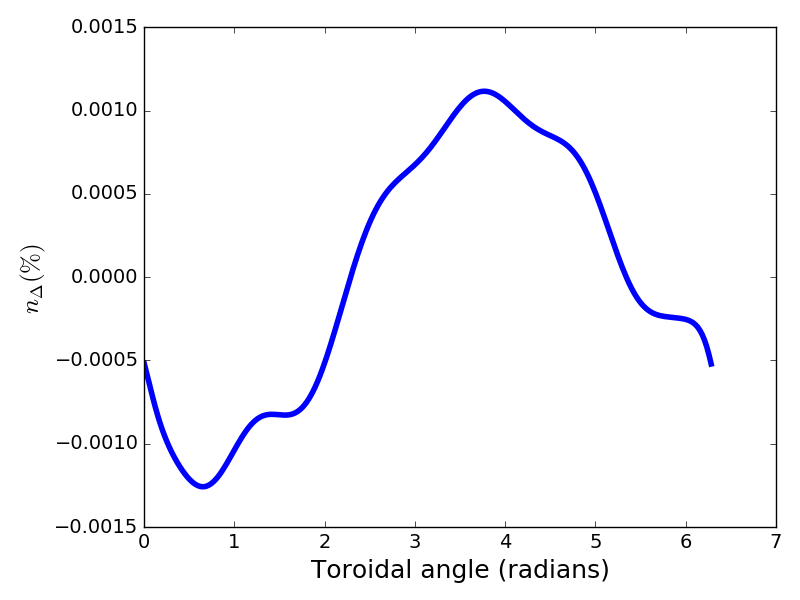}
        \caption{\label{fig:w7x_ndiff_profile} The profile of $n_\Delta$ as a percent difference from the mean along the field line in Figure~\ref{fig:ndiff_w7x}, indicating slightly non-uniform propagation.}
      \end{minipage}
    \end{minipage}
  \end{center}
\end{figure}

Following the implementation this curvature drive, filaments were initialized as in Figure~\ref{fig:initialconditions}, but with a background electron temperature of $T_{e0} = 250eV$.  Additionally, the parallel resolution was increased to 256 points (0.02 radians grid spacing) to accurately reproduce the fine structures -- presumably from local coil ripple -- seen in the Figure~\ref{fig:w7x_curvature}.  Figure~\ref{fig:ndiff_w7x} illustrates the predominantly radial propagation in this geometry as indicated by plotting $n_\Delta$, where positive values of $n_\Delta$ are red, and negative values are shown in blue.  Figure~\ref{fig:w7x_ndiff_profile} indicates the non-uniform nature of the propagation by plotting the $n_\Delta$ profile along a field line.  

\begin{figure}[h]
  \begin{center}
    \begin{minipage}{34pc}
      \begin{minipage}{16pc}
        \includegraphics[width=15pc]{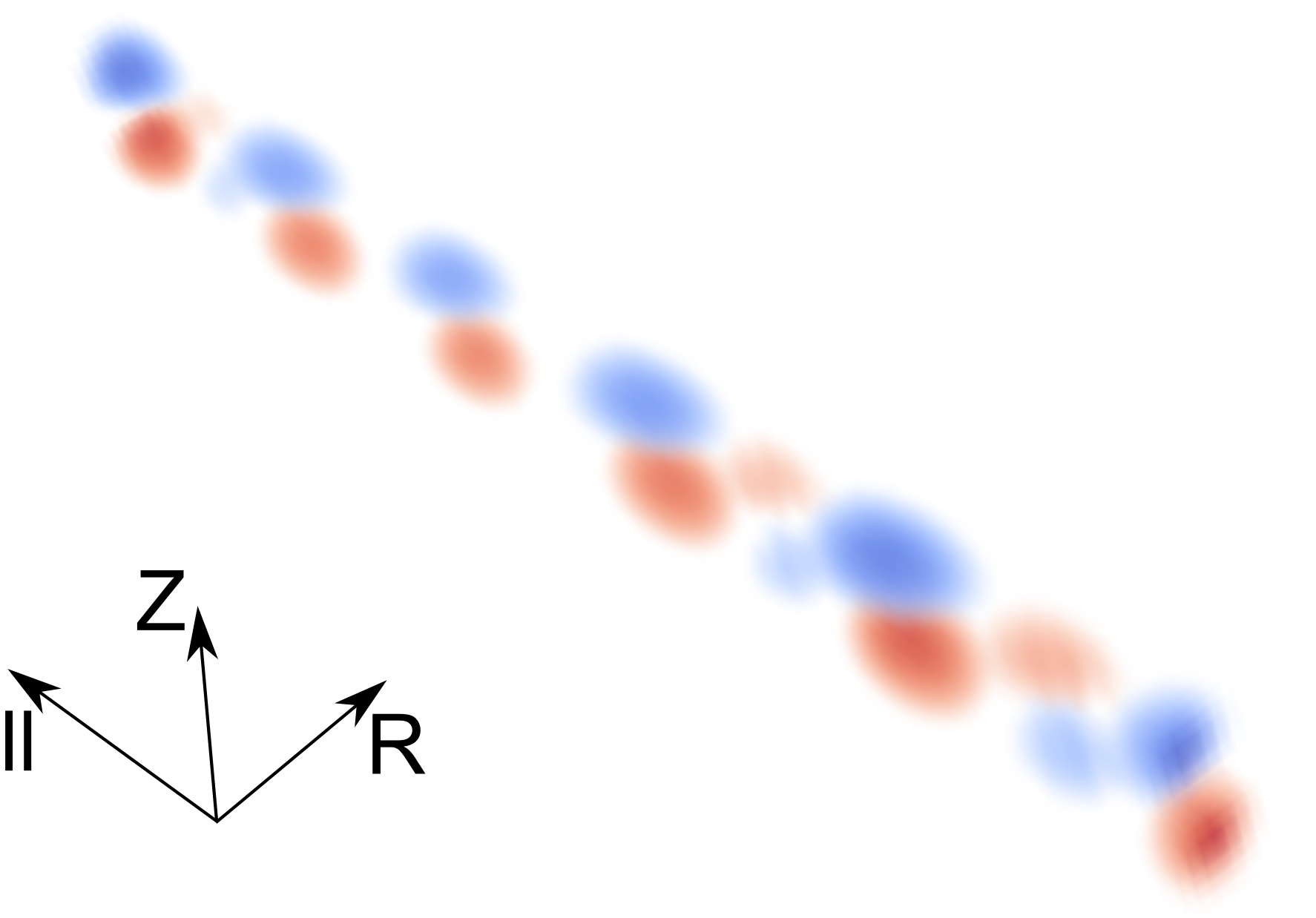}
        \caption{\label{fig:vorticity_early}Initial vorticity profile for a filament in Wendelstein 7-X, indicating strong non-uniformity.}
      \end{minipage}\hspace{2pc}%
      \begin{minipage}{16pc}
        \includegraphics[width=15pc]{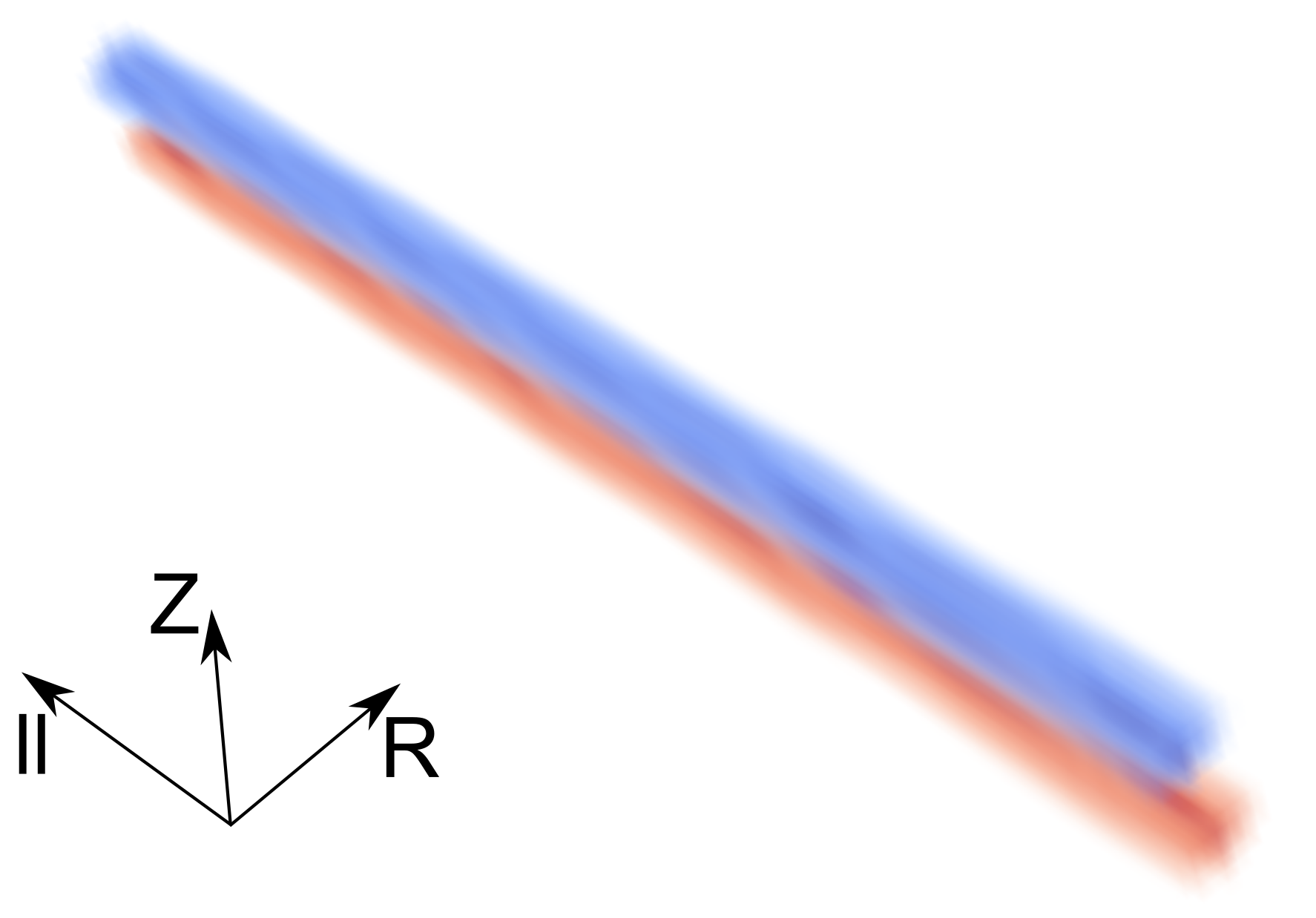}
        \caption{\label{fig:vorticity_late}Vorticity at a later time in a W7-X filament.  The initial non-uniformity is less visible, and the profile primarily reflects that of a radially-advecting filament.  Here again, positive values are red, negative values are blue.}
      \end{minipage}
    \end{minipage}
  \end{center}
\end{figure}

The non-uniform nature of the filament drive prominent in the initial stages of the filament evolution, as indicated by the vorticity at early time-steps, shown in Figure~\ref{fig:vorticity_early} which again uses a diverging color scheme such that positive values of vorticity are shown in red and negative values are shown in blue.  At a later time, Figure~\ref{fig:vorticity_late}, however, the field-line-averaged toroidal curvature dominates and the vorticity reverts to a similar state to that with a uniform toroidal curvature.

From this, one would expect that in the absence of a radial electric field, filaments in Wendelstein 7-X would propagate radially outward.  In stellarators, however, there exists a strong radial electric field.  A comparison of the poloidal propagation due to the radial electric field and the propagation due to the predominantly outward curvature drive should be a subject of future study, since the poloidal propagation could further reduce or otherwise complicate the radial propagation of filaments.

\section{Conclusions and Future Work}

An investigation of cold plasma filaments in periodic slab geometries with non-uniform curvature drive has been performed.  Firstly, it is shown that a curvature drive which changes sign does not shear a filament, unless the field-line-averaged drive is zero.  It is determined that the propagation of filaments predominantly follows the drive profile, but this effect can be reduced with higher temperatures.  The increased sound speed at higher temperatures favors the restoration to a uniform propagation profile.  It is also shown that propagation speed is dependent on the average curvature drive, but another factor must be involved as a stronger variation in curvature also increases the propagation.  Examining the aspects relevant in these scenarios, such as the parallel drive gradients, should be a subject of future work.

Simulations of filaments with the curvature drive along a field line in Wendelstein 7-X were also performed.  Despite a strongly non-uniform drive and initial vorticity profile, propagation was predominantly radially outward.  The simulations presented here, however, do not include the radial electric field.  The next step will be incorporating this effect and comparing the poloidal transport due to the radial electric field with the curvature drive to determine the expected propagation of a filament in this complicated magnetic geometry.

Additionally, these simulations can be extended to geometries with finite connection lengths, thereby providing relevance for edge and scrape-off-layer plasmas.  Following the development of the BSTING~\cite{Shanahan2018} project, simulations of full stellarator edge and scrape-off-layer geometries is also possible.  Therefore, verification of the work shown here in a broader geometry is foreseeable.   

\section{Acknowledgments}

The authors would like to acknowledge the work of the \boutxx~development team.

This work has been carried out within the framework of the EUROfusion Consortium and has received funding from the Euratom research and training programme 2014-2018 under grant agreement No 633053. The views and opinions expressed herein do not necessarily reflect those of the European Commission.

\section*{References}
\bibliography{BSTING}
\bibliographystyle{unsrt}


\end{document}